\documentclass[aip,apl,amsmath,amssymb,reprint]{revtex4-2}

\usepackage[english]{babel}
\usepackage[utf8]{inputenc}
\usepackage[colorinlistoftodos]{todonotes}
\usepackage[colorlinks,citecolor=blue,linkcolor=blue]{hyperref}
\usepackage{graphicx}	
\usepackage{bm}	
\usepackage{txfonts} 	
\usepackage[version = 4]{mhchem}     
\usepackage{siunitx}    
\usepackage{upgreek}
\usepackage{xcolor}
\usepackage{amsmath}
\usepackage[pagewise]{lineno}

\newcommand{\UIUC}{Department of Materials Science and Engineering and Materials Research Laboratory, The Grainger College of Engineering, University of Illinois at Urbana-Champaign, Illinois 61801, USA}
\newcommand{\NIU}{Department of Physics, Northern Illinois University, DeKalb, Illinois 60115, USA}
\newcommand{\Argonne}{Materials Science Division, Argonne National Laboratory, Lemont, Illinois 60439, USA}

\begin{document}
\setlength\columnsep{25pt}

\title{Micromagnetic simulations for magnetic multipoles}

\author{Myoung-Woo Yoo}
\email{mwyoo@illinois.edu}
\affiliation{\UIUC}
\author{Roland Winkler}
\affiliation{\NIU}
\affiliation{\Argonne}
\author{Axel Hoffmann}
\affiliation{\UIUC}

\date{\today}

\begin{abstract}
Cluster magnetic multipoles are order parameters that characterize the symmetry of spin arrangements in magnetic materials. In particular, high-order multipoles play a pivotal role in altermagnets and non-collinear antiferromagnets where they govern electrical and optical phenomena. While spatially non-uniform multipole textures have been observed on the micrometer scale, their behavior at mesoscopic lengths remains largely unexplored. Here, we introduce a comprehensive micromagnetic framework for vector-like cluster magnetic multipoles, enabling quantitative, spatially resolved analysis of non-uniform multipole systems. As a demonstration, we apply the framework to magnetic-octupole domain-wall motion in the non-collinear antiferromagnet \ce{Mn3Sn}. Our simulations capture key features of domain-wall dynamics, including profile deformation and the emergence of an effective inertial mass. This work provides a unified approach for investigating the mesoscopic dynamics of high-order cluster multipoles, and opens new avenues for understanding and engineering the physical properties of functional magnetic materials, such as altermagnets and non-collinear antiferromagnets, for advanced spintronic technologies.
\end{abstract}

\maketitle

\section{Introduction}
Antiferromagnets are magnetic materials in which the arrangement of magnetic dipole moments results in nearly zero net magnetization. Their vanishing stray fields, intrinsic stability, and ultrafast spin dynamics make them prime candidates for next-generation spintronics and high-performance information technologies~\cite{jungwirth_antiferromagnetic_2016, baltz_antiferromagnetic_2018, siddiqui_metallic_2020, chen_anomalous_2021, smejkal_emerging_2022, bai_altermagnetism_2024, song_altermagnets_2025, shim_spin-polarized_2025}. \par

Despite the absence of net magnetization, certain antiferromagnets, e.g., altermagnets and non-collinear antiferromagnets, exhibit spin-dependent electronic structures that give rise to unconventional electrical, optical, and thermal phenomena. Altermagnets are a particular class of antiferromagnets that maintain fully compensated spin order while exhibiting spin-split electronic bands originated from the interplay of crystal symmetry and spin configuration \cite{smejkal_emerging_2022, bai_altermagnetism_2024, song_altermagnets_2025}. Likewise, non-collinear antiferromagnets support large electrical and optical responses, such as the anomalous Hall effect and the magneto-optical Kerr effect, due to time-reversal-symmetry breaking by non-collinear spin order in kagome lattices that generates large Berry curvature~\cite{nakatsuji_large_2015, chen_anomalous_2021}. These phenomena are intimately linked to the symmetry of the spin arrangement and can be unified by cluster magnetic multipoles, which clarifies how symmetry dictates macroscopic observables in antiferromagnets~\cite{smejkal_anomalous_natrevmat_2022, suzuki_cluster_2017, chen_manipulating_2020, winkler_theory_2023, hayami_symmetry_2024, winkler_standard_2025}. \par

Prominent examples are the $D0_{19}$‑type non-collinear antiferromagnets including \ce{Mn3Sn} and \ce{Mn3Ge}, which possess a kagome crystal structure and an inverse-triangle spin configuration~\cite{kren_study_1975, tomiyoshi_magnetic_1982, brown_determination_1990, chen_anomalous_2014, kubler_non-collinear_2014}. These materials host Weyl nodes near the Fermi level and exhibit a non-vanishing Berry curvature, resulting in pronounced transport and optical responses despite their negligible net magnetization~\cite{nakatsuji_large_2015, ikhlas_large_2017, higo_large_2018}. In these antiferromagnets, a magnetic octupole moment serves as the primary order parameter which controls the sign and magnitude of transport and optical effects and can be manipulated by magnetic fields, electric currents, spin‑orbit and spin‑transfer torques, as well as thermal effects~\cite{brown_determination_1990, nakatsuji_large_2015, tsai_electrical_2020, xie_magnetization_2022, xie_efficient_2023, krishnaswamy_time-dependent_2022, pal_setting_2022, yoo_thermal_2024}. Recently, domain-wall-mediated octupole-state switching has been reported, enabling more efficient octupole control~\cite{wu_current-driven_2024}. \par

Just as ferromagnetic domain walls, cluster multipoles can form non-uniform multipole domain walls on mesoscopic scales~\cite{higo_large_2018,uchimura_observation_2022,higo_perpendicular_2022,yoo_thermal_2024,wu_current-driven_2024}. The dynamics of these textures are expected to differ from those of ferromagnetic domain walls owing to their compensated angular momentum, which suppresses conventional gyrotropic motion~\cite{liu_anomalous_2017, tsai_electrical_2020, takeuchi_chiral-spin_2021, higo_perpendicular_2022, yoon_handedness_2023}. \par

While analytical models have captured certain aspects of domain-wall motion, a comprehensive and spatially resolved description, which is essential both at mesoscopic scales and for inherently two-dimensional and three-dimensional domain-wall dynamics, requires numerical approaches \cite{yamane_dynamics_2019, wu_current-driven_2024, zhang_antiferromagnetic_2016, baltz_antiferromagnetic_2018, li_chiral_2019}. In ferromagnets, micromagnetic simulations bridge atomic‑scale physics and device‑scale phenomena by treating magnetization as a continuous field over finite volumes~\cite{brown_micromagnetics_1963, fidler_micromagnetic_2000, abert_micromagnetics_2019}. This approach uses a representative magnetization for a finite volume and associated energy terms based on continuum theory, allowing for efficient computation with a good balance between detail and computational feasibility, even though some atomic-scale details are lost. This approach has also been extended to ferrimagnets and collinear antiferromagnets, facilitating studies of mesoscopic domain structures, domain-wall motion, magnetization switching, and terahertz oscillations~\cite{puliafito_micromagnetic_2019,atxitia_landau-lifshitz-bloch_2012,suess_micromagnetic_2002}. However, a general micromagnetic framework for multipoles in non-collinear magnets remains insufficiently explored~\cite{yamane_dynamics_2019, nomoto_cluster_2020, wu_current-driven_2024}. With recent demonstrations of domain-wall motion in non-collinear antiferromagnets, investigating non-uniform magnetization dynamics on the micrometer scale has become both imperative and timely~\cite{wu_current-driven_2024}. \par

In this work, we develop a micromagnetic formalism for non-trivial magnetic systems that include cluster magnetic multipoles where the spin state within a unit cell can be described by a single representative vector. As a case study, we establish a micromagnetic model for the non-collinear antiferromagnet \ce{Mn3Sn}. We validate the model by performing micromagnetic simulations of a single octupole and one-dimensional \SI{60}{\degree} octupole domain-wall profiles, and comparing the results with those obtained from atomistic-spin model calculations. Building on this validation, we use the micromagnetic model to explore phenomena beyond the reach of atomistic-spin simulations. We study the field-driven motion of \SI{180}{\degree} domain walls composed of three consecutive \SI{60}{\degree} domain walls. Our results reveal characteristic wall deformation and emergent inertial mass, demonstrating that micromagnetics can capture key mesoscopic features of multipole dynamics. This framework provides a foundation for the quantitative study of mesoscopic multipolar textures in antiferromagnetic systems. \par

\section{Micromagnetic model for magnetic multipoles}

Consider a magnetic unit cell containing $n$ magnetic dipole moments, $\mathbf{m}_i$ = $m_\mathrm{s}$ ($\sin \theta_i$ $\cos \varphi_i$, $\sin \theta_i$ $\sin \varphi_i$, $\cos \theta_i$), at sub-lattices labeled $i = 1, \cdots, n$ ($n \geq 1$). The position of each site is $\mathbf{r}_i = \mathbf{r}_0+\Delta \mathbf{r}_i$, where $\mathbf{r}_0$ denotes the position of the unit cell and $\Delta \mathbf{r}_i$ is the displacement of sub-lattice~$i$. Representative examples are configurations of $\mathbf{m}_i$ where the collective spin state can be characterized by a cluster magnetic multipole moment \cite{suzuki_cluster_2017, nomoto_cluster_2020, winkler_theory_2023,winkler_standard_2025}. \par

\begin{figure}
\centering
\includegraphics[width = 7.4cm]{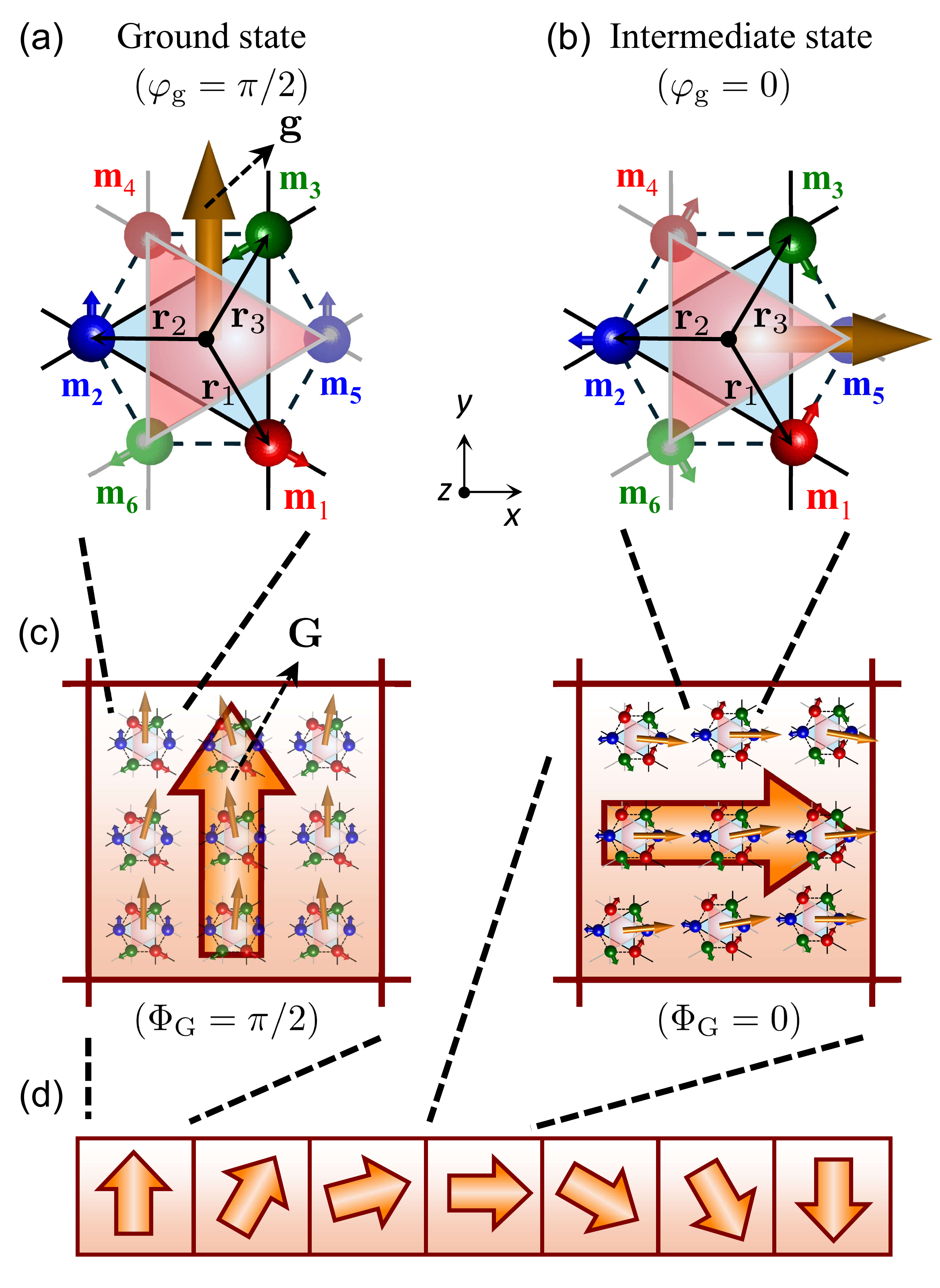}
\caption{\label{Fig:1}
    Micromagnetic model for non-collinear antiferromagnets. (a) Ground-state spin configuration of a \ce{Mn3Sn} unit cell. The red, blue, and green arrows represent magnetic dipole moments $\mathbf{m}_i$, on each sub-lattice, $i$, where Mn atoms are located. The black arrows show the position vectors, $\mathbf{r}_i$ of each sub-lattice. The bright-blue and pale-red planes mark two different kagome planes. Solid and dashed lines between sub-lattices represent intra‑layer and inter‑layer nearest‑neighbour exchange interactions, respectively. The large orange arrow is the characteristic vector $\mathbf{g}$ (Eq.~\ref{Eq:phiAngles}). In this case, $\mathbf{g}$ corresponds to the octupole moment. (b) Intermediate spin state in a \ce{Mn3Sn} unit cell at $\varphi$ = 0, naturally reached by rotating the individual magnetic moments, $\mathbf{m}_i$, from the configuration shown in (a). (c) A finite volume containing a sufficiently large number of magnetic unit cells. The large orange arrow shows the representative vector, $\mathbf{G}$, of the micromagnetic cell. (d) Example of a non-uniform multipole configuration represented by spatially varying $\mathbf{G}$.
}
\end{figure}

We assume that the relative orientations among the magnetic moments within a unit cell are well preserved so that they move rigidly together. In this case, they can jointly be characterized by three parameters representing the moments' orientation \cite{goldstein_classical_2002}. In certain cases the spin configuration is fully characterized by a single representative vector order parameter $\mathbf{g}$ = $g_0$ ($\sin \theta_\mathrm{g}$ $\cos \varphi_\mathrm{g}$, $\sin \theta_\mathrm{g}$ $\sin \varphi_\mathrm{g}$, $\cos \theta_\mathrm{g}$) requiring only two or one independent parameter. Examples are simple magnetic systems such as ferromagnets and collinear antiferromagnets, whose spin states are described by the magnetization vector and the N{\'e}el vector, respectively. Another class of systems that admits a single‑vector description comprises certain high‑rank magnetic‑multipole compounds with coplanar spin configurations, in which all magnetic dipole moments lie in a plane and their relative in‑plane angles remain fixed \cite{suzuki_cluster_2017}. In such cases, the collective spin state of each unit cell can be characterized by a vector $\mathbf{g}$ that specifies the in‑plane orientation of the cluster magnetic multipole moment [Fig.~\ref{Fig:1}]. \par

When the ensemble of spins in a unit cell moves rigidly together, a continuum micromagnetic model can be formulated by introducing a smoothly varying field $\mathbf{g} (\mathbf{r}, t)$. From this continuum description, we can derive the equation of motion for $\mathbf{g}$ with the micromagnetic energy terms. The equation of motion can be obtained from the Lagrangian density $\mathcal{L}$ and the Rayleigh dissipative function density $\mathcal{F}$ of the magnetic dipole moments,
\begin{equation} \label{Eq:Lagrangian_0}
    \begin{split}
        \mathcal{L} = & \sum_{i=1}^{n} \left\{ \dfrac{m_\mathrm{s}}{\gamma} \mathbf{A} (\hat{\mathbf{m}}_i) \cdot \dot{\hat{\mathbf{m}}}_i - \varepsilon_{i} \right\} \text{ and} \\        
        \mathcal{F} = & \sum_{i=1}^{n} \left\{ \alpha \dfrac{m_\mathrm{s}}{2 \gamma} \dot{\hat{\mathbf{m}}}_i^2 \right\} \text{.}
    \end{split}
\end{equation}
Here, $\hat{\mathbf{m}}_i$ = $\mathbf{m}_i/|\mathbf{m}_i|$, $m_\mathrm{s}$ is the magnitude of the individual magnetic dipole moment, $\gamma$ is a gyromagnetic ratio, $\mathbf{A}$ is the Berry-phase gauge potential, $\varepsilon_i$ is the potential energy of sub-lattice $i$, and $\alpha$ is a Gilbert damping constant~\cite{Döring+1948+373+379,alma9937635933902959}. By solving the Euler-Lagrangian equation, we obtain the equation of motion for $\mathbf{g}$ which serves as the counterpart to the Landau-Lifshitz equation for ferromagnets. These equations, with the associated energy terms, form the basis for micromagnetic simulation for non-collinear magnetic multipoles. \par

For numerical implementation, the continuum system is discretized into micromagnetic volumes $V_\mathrm{c}$, each containing a sufficient number of unit cells and the $\mathbf{g}$ vectors are nearly constant. Within each $V_\mathrm{c}$, a representative vector $\mathbf{G} (\mathbf{r}, t)$ is defined as the volume average of $\mathbf{g}$; $\mathbf{G} (\mathbf{r}, t)$ = $\frac{1}{V_\mathrm{c}} \int \frac{\mathbf{g} (\mathbf{r}, t)}{V_\mathrm{u}} d^3 r$, where $V_\mathrm{u}$ is the volume of a unit cell. The equation of motion for $\mathbf{G}$, derived from the continuum formulation, is then solved on this discretized grid with the computed energy terms, forming a micromagnetic simulation framework for multipole dynamics. \par

As an illustrative example, we develop a micromagnetic model for \ce{Mn3Sn}, a non-collinear antiferromagnet that exhibits a magnetic octupole moment. The magnetic unit cell of \ce{Mn3Sn} contains two kagome planes and six sub-lattices, $\mathbf{m}_i$, where $i = 1, \cdots, 6$ [Fig.~\ref{Fig:1}(a)]. The ground-state spin configuration is obtained from the Hamiltonian,
\begin{equation}
    \begin{split} \label{Eq:Hamiltonian}
        {\cal H} = &-J_1 \sum_{\langle ij \rangle_z} \hat{\mathbf{m}}_i \cdot \hat{\mathbf{m}}_j - J_2 \sum_{\langle ij \rangle_{xy}} \hat{\mathbf{m}}_i \cdot \hat{\mathbf{m}}_j \\
        &+ \sum_{\langle ij \rangle_{xy}} \mathbf{D} \cdot \left( \hat{\mathbf{m}}_i \times \hat{\mathbf{m}}_j \right) - \sum_i K \left( \hat{\mathbf{k}}_i \cdot \hat{\mathbf{m}}_i \right) ^2 \text{,}
    \end{split}
\end{equation}
where $\mathbf{m}_i$ is the normalized vector of magnetic dipole moments at each sub-lattice, $i$. $J_1$ and $J_2$ are isotropic exchange constants between inter- and intra-plane nearest neighbors, respectively. $\mathbf{D}=D\hat{\mathbf{z}}$ is the Dzyaloshinskii–Moriya interaction vector and $K$ is the single-ion anisotropy constant with local easy axis $\hat{\mathbf{k}}_i$~\cite{liu_anomalous_2017}. \par

The calculated ground state, illustrated in Fig.~\ref{Fig:1}(a), shows an anti-chiral triangular arrangement within the kagome plane, where the moments form angles close to \SI{120}{\degree}. Due to the local anisotropy, these moments slightly deviate, resulting in a small ferromagnetic component and emergent anisotropy~\cite{liu_anomalous_2017}. Another configuration, shown in Fig.\ref{Fig:1}(b), can be reached by rotating the moments, but it is unstable. The situation is reversed in \ce{Mn3Ge} where Fig.\ref{Fig:1}(b) is the ground state, and Fig.~\ref{Fig:1}(a) is unstable. Note that Figs.~\ref{Fig:1}(a) and (b) correspond to the $T^\gamma$ octupoles, according to the notation used in Ref.~\cite{suzuki_cluster_2017}. \par

We introduce a representative vector, $\mathbf{g}$ = $g_0$ ($\sin \theta_g \cos \varphi_g$, $\sin \theta_g \sin \varphi_g$, $\cos \theta_g$) with $\theta_g = \pi/2$ and the in-plane angles of the six moments are
\begin{equation} \label{Eq:phiAngles}
    \begin{split}
        \varphi_\text{1 or 4} &\approx (\pi-\varphi_g)-\dfrac{2\pi}{3} \text{,} \\
        \varphi_\text{2 or 5} &\approx (\pi-\varphi_g) \text{,} \\
        \varphi_\text{3 or 6} &\approx (\pi-\varphi_g)+\dfrac{2 \pi}{3} \text{,}
    \end{split}
\end{equation}
as in Fig.~\ref{Fig:1}(a). For clarity, small deviations from \SI{120}{\degree} are neglected in Eq.~\ref{Eq:phiAngles}. Here, we adopt a definition of $\mathbf{g}$ with a minus sign so that $\mathbf{g}$ aligns with both the magnetic octupole moment and the small net ferromagnetic moment of a \ce{Mn3Sn} unit cell. This choice simplifies the Zeeman energy term and makes the relation between the characteristic vector and the external magnetic field more intuitive. Note that choosing a different characteristic vector does not change our results [Supplementary Material]. \par

When the spin configuration varies slowly in space compared to the unit cell size, $\mathbf{g}$ can be promoted to a smoothly varying field $\mathbf{g} (\mathbf{r}, t)$. In this continuum limit, $\mathbf{g} (\mathbf{r}, t)$ serves as the multipole field used to formulate the micromagnetic formalism. \par

From Eqs.~\ref{Eq:Lagrangian_0} and \ref{Eq:phiAngles}, the Lagrangian density and the dissipative function density for $\mathbf{g} (\mathbf{r}, t)$ can be obtained. Solving the Euler-Lagrange equations, including spin-orbit torque terms [Supplementary Material], yields the equation of motion as
\begin{equation} \label{Eq:EqMotion_g}
    \begin{split}
        \dot{\mathbf{g}} (\mathbf{r}, t) = & -\gamma \mu_0 \mathbf{g} \times \mathbf{h}_{\mathrm{eff}} \quad \text{,}
    \end{split}
\end{equation}
where $\mathbf{h}_\mathrm{eff}$ is an effective field written as
\begin{equation} \label{Eq:heff}
    \begin{split}
        \mathbf{h}_{\mathrm{eff}} = \dfrac{1}{6 \mu_0 \alpha m_\mathrm{s} \rho} \left( \dfrac{\partial \mathcal{E}_\mathrm{u}}{\partial \varphi_\mathrm{g}} - \dfrac{\hbar p_z \theta_\mathrm{sh} j_\mathrm{hm}}{2 e d} \right) \hat{\mathbf{z}} \quad \text{,}
    \end{split}
\end{equation}
where $\mathcal{E}_\mathrm{u} = E_\mathrm{u}/V_\mathrm{u}$ is the energy density with the unit-cell energy $E_\mathrm{u}$. $\rho = 1/V_\mathrm{u}$ is the unit-cell density. Note that this equation of motion, Eqs.~\ref{Eq:EqMotion_g} and \ref{Eq:heff}, is fully consistent with previously reported equations of motion for the octupole moment in \ce{Mn3Sn} when a strong easy-plane constraint ($\theta_\mathrm{g} \approx \pi/2$) and the adiabatic condition ($\dot{\theta}_\mathrm{g} = 0$) are considered [Supplementary Material].~\cite{konakanchi_electrically_2025,rahman_strain_2025,he_magnetic_2024} \par

For numerical simulations, the continuum is discretized into micromagnetic volumes $V_\mathrm{c}$ that each contain a sufficient number of unit cells and $\mathbf{g} (\mathbf{r}, t)$ is nearly uniform within the cell. We then define the $\mathbf{g}$ density as a representative vector $\mathbf{G} (\mathbf{r}, t)$ = $\frac{1}{V_\mathrm{c}} \int \frac{\mathbf{g} (\mathbf{r}, t)}{V_\mathrm{u}} d^3 r$. Similar to conventional cases in \ce{Mn3Sn}, we can define the magnitude of the $\mathbf{g}$-vector, $g_0$, as the net ferromagnetic moment. Consequently, the magnitude of the $\mathbf{G}$ can be regarded as the effective magnetization associated with the octupole order. The equation of motion in Eq.~\ref{Eq:EqMotion_g} carries over to the discrete variables as 
\begin{equation} \label{Eq:EqMotion}
    \begin{split}
        \dot{\mathbf{G}} (\mathbf{r}, t) = & -\gamma \mu_0 \mathbf{G} \times \mathbf{H}_{\mathrm{eff}} \quad \text{,}
    \end{split}
\end{equation}
where 
\begin{equation} \label{Eq:Heff}
    \begin{split}
        \mathbf{H}_{\mathrm{eff}} = \dfrac{1}{6 \mu_0 \alpha m_\mathrm{s} \rho} \left( \dfrac{\partial \mathcal{E}_\mathrm{c}}{\partial \Phi_\mathrm{G}} - \dfrac{\hbar p_z \theta_\mathrm{sh} j_\mathrm{hm}}{2 e d} \right) \hat{\mathbf{z}} \quad \text{,}
    \end{split}
\end{equation}
where $\mathcal{E}_\mathrm{c}$ is a magnetic energy density in $V_\mathrm{c}$ and $\Phi_\mathrm{G}$ is the in-plane angle of $\mathbf{G}$ = $G_0$ ($\sin \Theta_\mathrm{G} \cos \Phi_\mathrm{G}$, $\sin \Theta_\mathrm{G} \sin \Phi_\mathrm{G}$, $\cos \Theta_\mathrm{G}$). Here, $\mathbf{G}$ has only an in-plane component because it is strongly confined to the kagome plane. Equation~\ref{Eq:EqMotion} is generally applicable to antiferromagnetic systems in which the magnetic moments are largely confined to the plane and the magnitude of $\mathbf{G}$ is approximately conserved. The limitations of this approximation are discussed further in the Supplementary Material. \par 

Next, we calculate the micromagnetic energy terms. The total magnetic energy density in the volume is given by $\mathcal{E}_\mathrm{c} = \mathcal{E}_\mathrm{zee} + \mathcal{E}_\mathrm{ani} + \mathcal{E}_\mathrm{exc}$, where $\mathcal{E}_\mathrm{zee}$, $\mathcal{E}_\mathrm{ani}$, $\mathcal{E}_\mathrm{exc}$ are the Zeeman, anisotropy, and exchange energy densities, respectively. The Dzyaloshinskii–Moriya interaction energy is included in $\mathcal{E}_\mathrm{exc}$, and demagnetization energy is neglected due to the tiny net magnetization. With $\mathbf{G}$ confined to the kagome plane, the Zeeman and anisotropy energy density terms become
\begin{equation} \label{Eq:Energies}
    \begin{split}
        \mathcal{E}_\mathrm{zee} = & -\mu_0 G_0 H_0 \sin \Theta_\mathrm{H} \cos \left( \Phi_\mathrm{G} - \Phi_\mathrm{H} \right) \text{,} \\
        \mathcal{E}_\mathrm{ani} = & K_\mathrm{6} \left( 1 + \cos 6 \Phi_\mathrm{G} \right) \text{,} \\
        \mathcal{E}_\mathrm{exc} = & A_\mathrm{ex, x} \left( \dfrac{\partial \Phi_\mathrm{G}}{\partial x} \right)^2 + A_\mathrm{ex, y} \left( \dfrac{\partial \Phi_G}{\partial y} \right)^2+ A_\mathrm{ex, z} \left( \dfrac{\partial \Phi_G}{\partial z} \right)^2 \text{,}
    \end{split}
\end{equation}
where $\mathbf{H}_\mathrm{ext}$ = $H_0$ ($\sin \Theta_\mathrm{H} \cos \Phi_\mathrm{H}$, $\sin \Theta_\mathrm{H} \sin \Phi_\mathrm{H}$, $\cos \Theta_\mathrm{H}$) is the external magnetic field~\cite{liu_anomalous_2017}. $A_{\mathrm{ex},x}$, $A_{\mathrm{ex},y}$, $A_{\mathrm{ex},z}$ are the exchange stiffness constants in the $x$-, $y$-, $z$-directions, respectively, which can be expressed as
\begin{equation} \label{Eq:Aex_xyz}
    \begin{split}
        A_{\mathrm{ex},x} = A_{\mathrm{ex},y} &= - \dfrac{J_1+3 J_2+3 D \sqrt{3}}{8 \sqrt{2} a_0} \quad \text{and} \\
        A_{\mathrm{ex},z} &= - \dfrac{J_1 }{2 \sqrt{2} a_0}  \text{.}
    \end{split}
\end{equation}
where $a_0 \approx 0.28$~nm is a distance between the nearest sub-lattices [Supplementary Material]. Note that while the slight rotation of $\mathbf{m}_i$ in the non-uniform texture also affects the anisotropy energy $\mathcal{E}_\mathrm{ani}$, the impact is negligible in \ce{Mn3Sn} because the exchange constants $J_1$ and $J_2$ are significantly larger than the anisotropy constant, $K_6$. \par

\ce{Mn3Sn} ideally exhibits six-fold in-plane anisotropy characterized by $K_6$, but thin films may also show an additional two-fold in-plane anisotropy due to tensile strain~\cite{ikhlas_piezomagnetic_2022, higo_perpendicular_2022, yoon_handedness_2023}. In such case, the anisotropy energy density in Eq.~\ref{Eq:Energies} may include an additional term, $K_\mathrm{2} \cos^2 \Phi_G$. \par

\section{Micromagnetic Simulations}

\subsection{Single Octupole Magnetization Dynamics}

\begin{figure}
\centering
\includegraphics[width = 8.0cm]{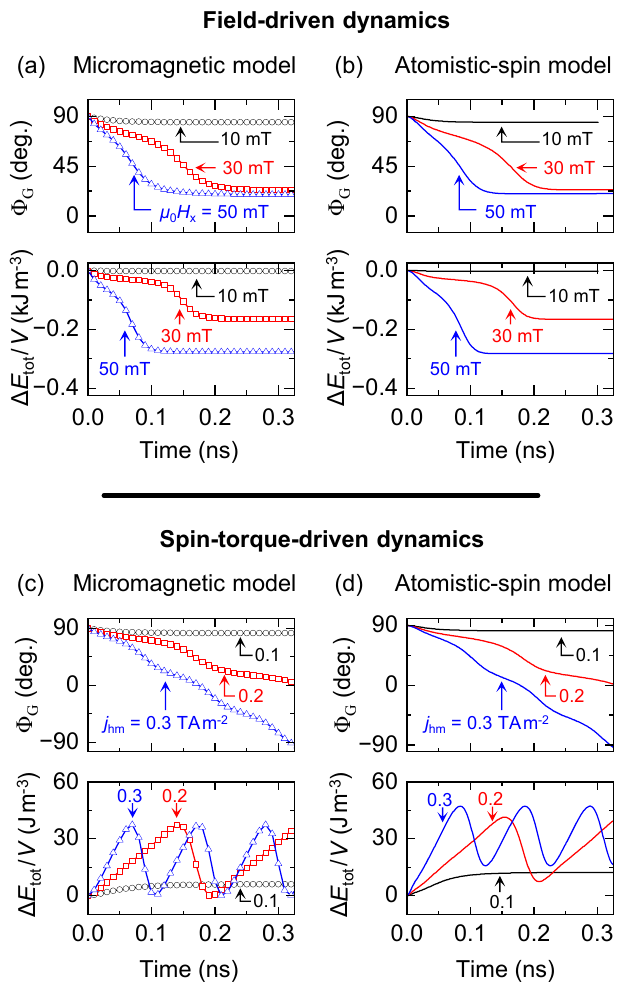}
\caption{\label{Fig:2}
    Comparison between the micromagnetic and the atomistic-spin simulations for an octupole magnetization dynamics (a) Time evolution of the octupole angle, $\Phi_\mathrm{G}$, (top) and the total magnetic energy change, $\Delta E_\mathrm{tot}$, (bottom) calculated with the micromagnetic model under different $+x$-directional magnetic fields, $H_x$. $\Delta E_\mathrm{tot}$ is normalized by the volume, $V$. (b) Same as (a), but calculated with the atomistic-spin model. (c) $\Phi_\mathrm{G}$ (top) and $\Delta E_\mathrm{tot}$ (bottom) calculated using the micromagnetic model under a spin-orbit torque at current density $j_\mathrm{hm}$ in an adjacent heavy-metal layer. The spin polarization $\mathbf{p}$ is in the $+z$-direction. (d) Same as (c), but obtained with the atomistic-spin model.
}
\end{figure}

Using the micromagnetic model, we numerically calculated the dynamics of $\mathbf{G}$ in \ce{Mn3Sn}. In this work, we used parameters: $J_1$ = $J_2$ = $-2.803$ meV, $D$ = $-0.635$ meV, $K$ = 0.187 meV, and $m_\mathrm{s} = 3 \mu_\mathrm{B}$, where $\mu_\mathrm{B}$ is the Bohr magneton. From numerical calculations, we obtained $G_0 = 6.1 \times 10^3$ \si{\ampere \meter^{-1}} and $K_6 = 37.5$ \si{\joule \meter^{-3}}. We use the numerically obtained values in this work, while these parameters can also be calculated analytically. [see Supplementary Material]~\cite{liu_anomalous_2017}. \par

In the first simulation, we focused on field-driven octupole dynamics. Initially, the octupole was oriented in the $+y$-direction, and a magnetic field parallel to the $+x$-direction was applied to induce the octupole motion. The micromagnetic cell was 10 $\times$ 10 $\times$ 10 \si{\nano\meter^3}, and periodic boundary conditions were applied in all three directions. We obtained the time evolution of $\Phi_\mathrm{G}$ and the change in total magnetic energy, $\Delta E_\mathrm{tot}$, by solving Eqs.~\ref{Eq:EqMotion} and \ref{Eq:Energies} with a Runge-Kutta method. The results are shown in Fig.~\ref{Fig:2}(a). For a small magnetic field, $\mu_0 H_x$ = 10 mT, the octupole remains near $\Phi_\mathrm{G}$ = \SI{90}{\degree}, because it cannot overcome the energy barrier. For larger fields, $\mu_0 H_x$ = 30 or 50 mT, the octupole rotates and stabilizes near the other ground state at $\Phi_\mathrm{G}$ = \SI{30}{\degree}. During the motion, $E_\mathrm{tot}$ decreases due to the reduction in Zeeman energy. We also conducted the same simulation using the atomistic-spin model as shown in Fig.~\ref{Fig:2}(b), and obtained nearly identical results. \par

In the second simulation, we calculated the octupole rotation induced by spin-orbit torque at a current density $j_\mathrm{hm}$ in an adjacent heavy-metal layer. For this simulation, we used a spin-Hall angle of $\theta_\mathrm{sh}$ = 0.1  and a \ce{Mn3Sn}-film thickness $d$ = 40 nm. The spin polarization $\mathbf{p}$ was parallel to the $+z$-direction. The results are shown in Fig.~\ref{Fig:2}(c). When $j_\mathrm{hm}$ is larger than the critical value, $j_\mathrm{hm}$ = 0.2 and $0.3 \times 10^{12}$ \si{\ampere\meter^{-2}}, the octupole begins continuous rotation, and $\Delta E_\mathrm{tot}$ oscillates due to the six-fold anisotropy, while the magnetization does not rotate when $j_\mathrm{hm}$ is small, $j_\mathrm{hm}$ = $0.1 \times 10^{12}$ \si{\ampere\meter^{-2}}. We also obtained a similar behavior from the atomistic-spin model as depicted in Fig.~\ref{Fig:2}(d). Note that the atomistic-spin model shows a slight initial increase in $\Delta E_\mathrm{tot}$ due to a small spin rearrangement induced by the spin-orbit torque. After 0.05~ns, however, the subsequent dynamics of $\Delta E_\mathrm{tot}$ closely match those of the micromagnetic model. \par

The consistency between the results from the micromagnetic and the atomistic-spin models, as shown in Fig.~\ref{Fig:2} confirms the agreement between the two models in the uniform state. \par

\subsection{60-degree Domain Walls}

\begin{figure}
\centering
\includegraphics[width = 8.0cm]{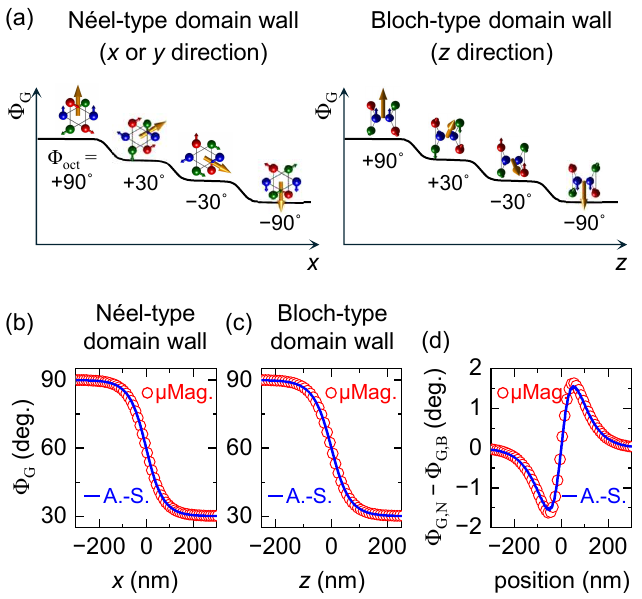}
\caption{\label{Fig:3}
    Profiles of domain walls in \ce{Mn3Sn}. (a) Schematic of N{\'e}el-type (left) and Bloch-type (right) domain walls propagating along the $x$ and $z$ directions, respectively. Orange arrows indicate the octupole-vector orientation. A \SI{180}{\degree} domain wall consists of three consecutive \SI{60}{\degree} domain walls. (b) Octupole-angle profile $\Phi_\mathrm{G}$ for a N{\'e}el-type \SI{60}{\degree} domain wall along the $x$ direction, calculated with the micromagnetic model ($\mu$Mag., symbols) and the atomistic-spin model (A.-S., solid line). (c) Same as (b), but for a Bloch-type domain wall along the $z$ direction (d) Difference in $\Phi_\mathrm{G}$ between (b) and (c), obtained from the micromagnetic model ($\mu$Mag., symbols) and the atomistic-spin model (A.-S., solid line).
}
\end{figure}

We investigated non-uniform octupole configuration, one-dimensional magnetic domain walls, using the micromagnetic model. In \ce{Mn3Sn}, the kagome-plane orientation determines the type of domain wall \cite{wu_current-driven_2024}. For example, N{\'e}el-type walls are formed when the propagation direction is along the $x$ or $y$ directions, whereas Bloch-type walls appear when the propagation is along the $z$ direction [Fig. \ref{Fig:3}(a)]. In addition, a \SI{180}{\degree} domain wall in \ce{Mn3Sn} can have a staircase-like structure consisting of three \SI{60}{\degree} domain walls due to the six-fold anisotropy \cite{sugimoto_electrical_2020}. In this study, we numerically calculated the octupole profile of the \SI{60}{\degree} domain wall, where $\Phi_\mathrm{G}$ varies from +\SI{+90}{\degree} to +\SI{+30}{\degree}. Hereafter, we refer to this domain wall as a [+\SI{90}{\degree}, +\SI{30}{\degree}] domain wall. \par

The profiles of the N{\'e}el- and Bloch-type domain walls are plotted in Figs.~\ref{Fig:3}(b) and \ref{Fig:3}(c) (symbols), respectively, obtained from the micromagnetic model. We additionally plotted the profiles simulated from the atomistic-spin model using solid lines, which closely match the results from the micromagnetic model. Note that the micromagnetic model is computationally far more efficient than the atomistic-spin model [see Supplementary Material]. \par

\begin{figure}
\centering
\includegraphics[width = 8.0cm]{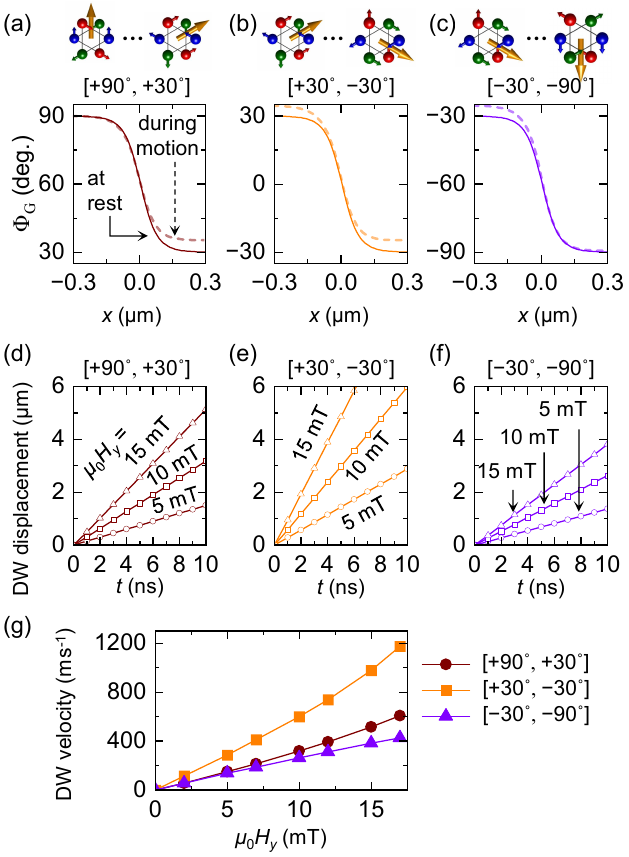}
\caption{\label{Fig:4}
    Motion of \SI{60}{\degree} N{\'e}el-type domain walls (DWs). (a)-(c) Octupole-angle profiles $\Phi_\mathrm{G}$ at rest (solid lines) and during motion (dashed lines) under $\mu_0 H_y$ = 10 mT for the [+\SI{+90}{\degree}, +\SI{+30}{\degree}], [+\SI{+30}{\degree}, \SI{-30}{\degree}], and [\SI{-30}{\degree}, \SI{-90}{\degree}] domain walls, respectively. (d)-(f) Time evolution of the domain-wall positions under different $H_y$ for the three \SI{60}{\degree} domain walls. (g) Domain-wall velocities as a function of $H_y$.
}
\end{figure}

The \SI{60}{\degree}-domain-wall profiles in \ce{Mn3Sn} can be expressed analytically as $\Phi_\mathrm{G} = -(1/3) \arctan \{ \sinh (\pi x / \delta_\mathrm{w} ) \}+\pi/3$, where the domain-wall width is $\delta_\mathrm{w} = (\pi \sqrt{A_{ex}}) / (3 \sqrt{K_6})$ [see Supplementary Material]. In our system, $\delta_\mathrm{w}$ = 145 nm and 128 nm for the N{\'e}el- and Bloch-type domain walls, respectively. Consequently, the two types show a slight difference as shown in Fig.~\ref{Fig:3}(d). These results demonstrate that the micromagnetic model can accurately describe the non-uniform octupole configurations. \par

Next, we studied the domain-wall motion by applying a magnetic field $H_y$. Once the field is applied to a N{\'e}el-type domain wall, it begins to move with a finite velocity, $v_\mathrm{dw}$. The domain-wall motion involves shape deformation due to changes in the octupole angles of two domains. In Figs.~\ref{Fig:4}(a)-\ref{Fig:4}(c), we show the profiles of three \SI{60}{\degree} domain walls, both at rest (solid lines) and during the steady-state motion (dashed lines), calculated with the micromagnetic model. Note that we could not compute the domain-wall motion with the atomistic-spin model because of our computational resource limitations. \par

The time evolution of the wall position and the velocities are plotted in Figs.~\ref{Fig:4}(d)-~\ref{Fig:4}(f) and Fig.~\ref{Fig:4}(g), respectively. The velocity increases monotonically with increasing $H_y$ for all domain walls, but the velocities differ for each one. The [+\SI{+30}{\degree}, \SI{-30}{\degree}] wall moves at roughly twice the speed of the others. At lower fields the velocities of the [+\SI{+90}{\degree}, +\SI{+30}{\degree}] and [\SI{-30}{\degree}, \SI{-90}{\degree}] walls are nearly identical, but the former becomes gradually faster at higher fields. These results imply that a \SI{180}{\degree} domain wall in \ce{Mn3Sn} can undergo complex deformation during motion because its constituent \SI{60}{\degree} domain walls move at different speeds. \par

Before studying the dynamics of a \SI{180}{\degree} domain wall, we briefly compare domain-wall motion in \ce{Mn3Sn} with that in conventional ferromagnets. At low magnetic fields, the two systems exhibit qualitatively similar field-driven dynamics, with the domain-wall velocity approximately proportional to the field magnitude. At sufficiently high magnetic fields, \ce{Mn3Sn} may also exhibit Walker-breakdown-like dynamics similar to those of ferromagnetic domain walls, due to the induced out-of-plane magnetic component, $\hat{m}_z$. In the present simulations, however, we do not observe such an instability because we employ an easy-plane reduced model for the micromagnetic calculations, in which the out-of-plane degree of freedom is assumed to remain small. Quantitative estimates based on the equations of motion that include the out-of-plane dynamics show that the induced out-of-plane magnetic component is only $\hat{m}_z \approx 2 \times 10^{-4}$ even at $v_\mathrm{dw} = 1.2$~\unit{\kilo\meter\per\second}, the highest domain-wall velocity shown in Fig.~\ref{Fig:4}(g). If the out-of-plane degree of freedom is retained explicitly, a Walker-breakdown-like behavior may, in principle, arise in this system. Our estimates indicate that such a regime would require canting angles of several degrees, well beyond those reached in the present simulations [see Supplementary Material]. \par

\subsection{180-degree Domain Walls}

We calculated the profile and motion of a \SI{180}{\degree} domain wall using the micromagnetic model. Due to the six-fold symmetric anisotropy in \ce{Mn3Sn}, the \SI{180}{\degree} domain wall is expected to exhibit a staircase-like structure composed of three \SI{60}{\degree} domain walls~\cite{sugimoto_electrical_2020,wu_magnetic_2023}, although such a step-like internal structure has not yet been directly resolved experimentally.~\cite{higo_large_2018,uchimura_observation_2022,wu_magnetic_2023,reichlova_imaging_2019,li_nanoscale_2023,tsukamoto_observation_2025} Note that there is no unique optimal spacing between the \SI{60}{\degree} domain walls without considering magnetostatic energy or an additional anisotropy. To address this in our simulations, we introduced an extra uniaxial anisotropy in the $y$ direction, i.e., $E_\mathrm{ani}$ = $V_\mathrm{c}$ $\left\{ K_\mathrm{6} (1+ \cos 6 \Phi_\mathrm{G}) + K_2 (1+ \cos 2 \Phi_\mathrm{G}) \right\}$. We used $K_2$ = \SI{10}{\joule\meter^{-3}} which is about \SI{30}{\percent} of $K_6$ = \SI{37.5}{\joule\meter^{-3}}. It has been reported that such uniaxial anisotropy can be introduced by in-plane tensile strain in \ce{Mn3Sn}~\cite{ikhlas_piezomagnetic_2022, higo_perpendicular_2022, yoon_handedness_2023}. In principle, a characteristic spacing between the \SI{60}{\degree} domain walls could also be achieved by including dipole–dipole interactions. \par

\begin{figure}
\centering
\includegraphics[width = 8.0cm]{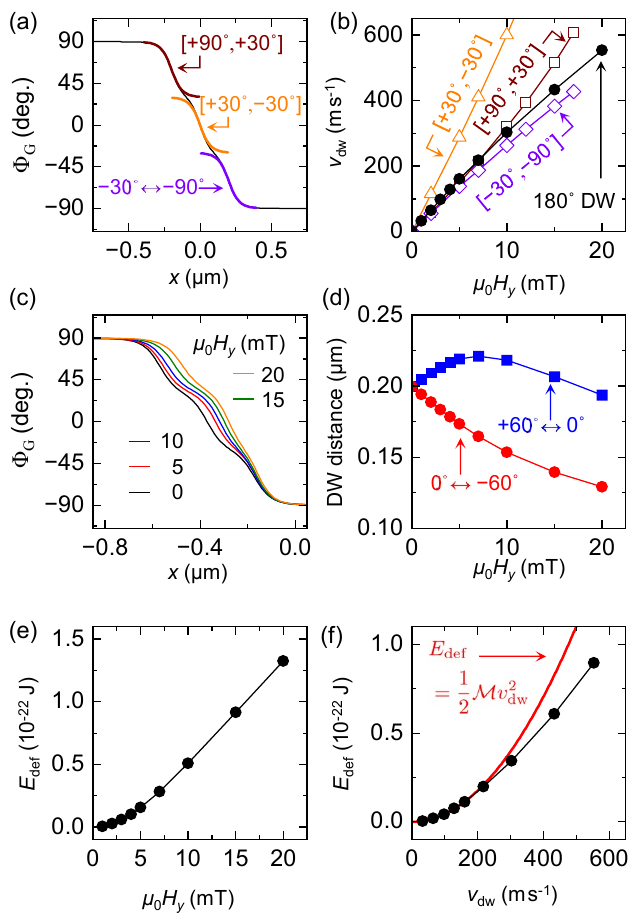}
\caption{\label{Fig:5}
    \SI{180}{\degree} domain-wall profiles and dynamics. (a) Profiles of a \SI{180}{\degree} domain wall from the micromagnetic model (black curve). Brown, orange, and violet curves are three \SI{60}{\degree} domain walls from Figs.~\ref{Fig:4}(a)-(c). (b) Steady-state velocity $v_\mathrm{dw}$ of the \SI{180}{\degree} domain wall (black closed circle) and of the three constituent\SI{60}{\degree} walls, [+\SI{+90}{\degree}, +\SI{+30}{\degree}] (brown open square), [+\SI{+30}{\degree}, \SI{-30}{\degree}] (orange open triangle), and [\SI{-30}{\degree}, \SI{-90}{\degree}] (violet open diamond), under $H_y$. (c) \SI{180}{\degree} domain-wall profiles during steady-state motion for different $\mu_0 H_y$. (d) Distances between $\Phi_\mathrm{G}$ = +\SI{60}{\degree} and \SI{0}{\degree} positions (red), and $\Phi_\mathrm{G}$ = +\SI{0}{\degree} and \SI{-60}{\degree} positions (blue) as a function of $\mu_0 H_y$. (e) Deformation energy $E_\mathrm{def}$ as a function of $H_y$. (f) $ E_\mathrm{def}$ as a function of $v_\mathrm{dw}$ (black circles and line). The red line shows a quadratic fit at low $v_\mathrm{dw}$. ${\mathcal M}$ denotes the effective inertial mass extracted from the fit.
}
\end{figure}

Figure.~\ref{Fig:5}(a) shows the calculated profile of a \SI{180}{\degree} N{\'e}el-type domain wall. The slopes and shapes of each step are in good agreement with those of the individual \SI{60}{\degree} domain walls as shown by colored lines in Fig.~\ref{Fig:5}(a). Note that the total width of the \SI{180}{\degree} domain wall decreases with increasing $K_2$ [see Supplemental Material]. \par

Next, we induced the motion of the \SI{180}{\degree} domain wall by applying a magnetic field in the $+y$ direction $H_y$. As shown in Fig.~\ref{Fig:5}(b) (black closed circle), the domain-wall velocity $v_\mathrm{dw}$ increases with increasing $H_y$, although its slope decreases slightly similar to the [\SI{-30}{\degree}, \SI{-90}{\degree}] domain wall shown in Fig.~\ref{Fig:4}(g). $v_\mathrm{dw}$ of the \SI{180}{\degree} domain wall is faster than the [\SI{-30}{\degree}, \SI{-90}{\degree}] domain wall, but slower than the [+\SI{+30}{\degree}, \SI{-30}{\degree}] and the [+\SI{+90}{\degree}, +\SI{+30}{\degree}] walls. This result shows that the \SI{180}{\degree}-domain-wall motion is determined by mutual interaction between the three \SI{60}{\degree} domain walls. We also obtained the relation between $v_\mathrm{dw}$ and $H_y$ for the $z$-directional Bloch domain wall as provided in the Supplementary Material. \par

Because the different \SI{60}{\degree} domain walls move with different velocities under $H_y$, the \SI{180}{\degree} domain wall deforms during motion. In Fig.~\ref{Fig:5}(c), we plot the $\Phi_\mathrm{G}$ profiles during steady-state motion for several $\mu_0 H_y$. At the ground state, the three walls are evenly spaced. However, during motion, the total width of the domain wall shrinks. In particular, the spacing between the [+\SI{+30}{\degree}, \SI{-30}{\degree}] and [\SI{-30}{\degree}, \SI{-90}{\degree}] walls decreases as $H_y$ increases. To quantify this, we plotted the distances between the wall centers, i.e., $\Phi_\mathrm{G}$ = \SI{60}{\degree}, \SI{0}{\degree}, and \SI{-60}{\degree}, in Fig.~\ref{Fig:5}(d) as a function of $\mu_0 H_y$. For $\mu_0 H_y <$ 7 mT, the distance between \SI{0}{\degree} and \SI{-60}{\degree} positions and +\SI{+60}{\degree} and \SI{0}{\degree} positions monotonically decreases and increases with increasing $H_y$, respectively, reflecting the different velocities. However, the distance between +\SI{+60}{\degree} and \SI{0}{\degree} begins to decrease from $H_y \approx$ 7~mT, to reduce the anisotropy energy. \par

The change in the slope sign can be qualitatively understood in terms of the motion of individual \SI{60}{\degree} walls. Because the central \SI{60}{\degree} wall, [+\SI{+30}{\degree}, \SI{-30}{\degree}], is much faster than other two, the distances between left and central walls [blue line and squares in Fig.~\ref{Fig:4}(d)] and right and central walls [red line and dots in Fig.~\ref{Fig:4}(d)] initially increase and decrease, depending on their relative velocities. However, these separations cannot grow indefinitely, as variations in wall distances increase the magnetic energy and induce restoring forces between the walls. As a result, the competition between differential wall velocities and inter-wall interaction leads to a non-monotonic dependence of the distance between left and central walls, producing a maximum separation near $H_y \approx$ 7~mT. \par

Deformation during the motion increases the magnetic energy of the domain wall, as shown in Fig.~\ref{Fig:5}(e) similar to ferromagnetic domain walls~\cite{alma9937635933902959}. In ferromagnets, the domain-wall angle changes with increasing $v_\mathrm{dw}$ and the deformation increases the domain-wall energy, $E_\mathrm{def}$. When $v_\mathrm{dw}$ is slow, $E_\mathrm{def} \propto v_\mathrm{dw}^2$ approximately, and an effective inertial mass of the domain wall is defined as ${\mathcal M} = 2 E_\mathrm{def} / v_\mathrm{dw}^2$. \par

To obtain an effective inertial mass, ${\mathcal M}$, of the \SI{180}{\degree} domain wall in \ce{Mn3Sn}, we plotted the deformation energy, $E_\mathrm{def}$, as a function of domain-wall velocity, $v_\mathrm{dw}$, in Fig.~\ref{Fig:5}(f). At low $v_\mathrm{dw}$, the relation follows a quadratic dependence, yielding the domain-wall mass, ${\mathcal M} \approx 8.8 \times 10^{-28}$~kg. A Bloch-type domain wall has a mass roughly twice as large, ${\mathcal M} \approx 1.7 \times 10^{-27}$~kg [see Supplementary Material]. The mass per unit cross-sectional area for the N{\'e}el- and Bloch-type domain walls are $8.8 \times 10^{-12}$ and $1.7 \times 10^{-11}$~\si{\kilo\gram~\meter^{-2}} which are comparable to the mass density of a ferromagnetic domain wall in Permalloy, $3 \times 10^{-11}$~\si{\kilo\gram~\meter^{-2}}~\cite{rhensius_imaging_2010}. The finite inertial mass implies that the octupole domain walls can exhibit inertial motion [see Supplementary Material], and the mass could be experimentally measurable, as in ferromagnets~\cite{saitoh_current-induced_2004, rhensius_imaging_2010, buttner_dynamics_2015}. Moreover, such inertial spin dynamics can significantly influence spin dynamics and spin switching as predicted in collinear antiferromagnetic systems~\cite{kimel_inertia-driven_2009}.

%
\section{Conclusion}
We have proposed a comprehensive micromagnetic formalism for non-uniform magnetic-multipole dynamics at the micrometer scale in non-collinear magnetic systems, analogous to conventional micromagnetic models for ferromagnets. As a demonstration, we numerically investigated the octupole domain-wall dynamics in the non-collinear antiferromagnet \ce{Mn3Sn} using this framework. The model successfully captures the key features of both \SI{60}{\degree} and \SI{180}{\degree} octupole domain-wall dynamics, which are challenging to predict with analytical or atomistic-spin models. Our simulations reveal that field-driven domain-wall motion induces profile deformations, leading to an increase in magnetic energy. By analyzing this deformation-induced energy increase, we quantified the effective inertial mass of the domain wall in a non-collinear antiferromagnet. This study establishes a general framework for exploring the mesoscopic dynamics of higher-order magnetic multipoles in antiferromagnets, crucial for understanding and engineering non-trivial magnetic materials, such as altermagnets and non-collinear antiferromagnets, for advanced spintronic technologies.
%

%
\section{Supplementary Material}

See the supplementary material for detailed derivations of the equation of motion for the magnetic multipole vector, analytical expressions for the octupole magnetization and six-fold anisotropy, and additional micromagnetic simulation results.

\section{acknowledgments}
This research is supported by the NSF through the University of Illinois Urbana-Champaign Materials Research Science and Engineering Center Grant No. DMR-1720633 and is carried out in part in the Materials Research Laboratory Central Research Facilities, University of Illinois.  Work at Argonne was supported by DOE BES under Contract No.\ DE-AC02-06CH11357.

\section{AUTHOR DECLARATIONS}
\subsection{Conflict of Interest}
The authors have no conflicts to disclose
\subsection{Author Contributions}
Myoung-Woo Yoo: Conceptualization (equal); Data Curation (lead); Formal Analysis (lead); Investigation (lead); Methodology (equal); Software (lead); Visualization (lead); Writing/Original Draft Preparation (equal). Roland Winkler: Conceptualization (equal); Methodology (equal); Writing/Original Draft Preparation (equal). Axel Hoffmann: Conceptualization (equal); Funding Acquisition (lead); Project Administration (lead); Supervision (lead); Writing/Original Draft Preparation (equal).
\section{DATA AVAILABILITY}
The data that support the findings of this study are available from the corresponding author upon reasonable request.

\bibliography{references}{}
\end{document}